\documentclass[12pt,preprint]{aastex}

\usepackage{epsfig}
\usepackage{amsmath,amssymb}

\shorttitle{Reconnection in Disk Coronae}
\shortauthors{Goodman and Uzdensky}

\newcommand{\unit}[1]{\,\textrm{#1}}

\newcommand{\dSP}{\delta_{\rm SP}}
\newcommand{\di}{{d}_i}
\newcommand{\RS}{R_{\rm S}}
\newcommand{\sigmaT}{\sigma_{\rm T}}
\newcommand{\VA}{V_{\rm A}}
\newcommand{\VS}{V_{S}}

\begin{document}

\title{Reconnection in Marginally Collisionless Accretion Disk Coronae}


\author{J. Goodman and D. Uzdensky}
\affil{Princeton University Observatory, Princeton, NJ 08544}

\begin{abstract}
  We point out that a conventional construction placed upon observations 
  of accreting black holes, in which their nonthermal X-ray spectra are 
  produced by inverse comptonization in a coronal plasma, suggests that 
  the plasma is marginally collisionless.  Recent developments in
  plasma physics indicate that fast reconnection takes place only in 
  collisionless plasmas. As has recently been suggested for the Sun's 
  corona, such marginal states may result from a combination of
  energy balance and the requirements of fast magnetic reconnection.
\end{abstract}



\date{\today}



\section{Introduction}
\label{sec-intro}

The spectral states of X-ray binaries are distinguished by the
relative strength of a soft quasi-thermal component and a harder
power-law component.  The soft component is attributed to an optically 
thick accretion disk, while the power law is usually ascribed to a hotter, 
geometrically thicker, and optically thinner distribution of plasma called 
the corona [see \cite{Zdziarski_Gierlinski04} for a recent review].  
In some models the corona lies directly above the disk 
[the ``sandwich model'', \cite{Liang_Price77}; see also 
\cite{Uzdensky_Goodman08} for more complete references], 
while in others the corona lies inside the inner edge of a radially 
truncated disk \cite[e.g.,][]{Thorne_Price75,Shapiro_etal76}.  
During hard states at least, more energy is dissipated in the corona 
than in the disk.  The coronal emission mechanism is usually
identified as inverse
comptonization of soft photons from the disk by hot, thermal coronal electrons
\citep{Eardley_Lightman76},
although synchrotron-self-Compton emission from relativistic electrons has
also been proposed \citep[e.g.][]{Band_Grindlay86,Field_Rogers93}.
To explain the observed power-law spectra, the Compton parameter
$y\equiv (4kT_e/m_e c^2)\, {\rm max}(\tau,\tau^2)$ must be of order unity.  
Here $\tau\equiv n_e\sigmaT H$ is the optical depth of the corona to 
electron scattering, $H$ its vertical scale height, and $T_e$ its electron
temperature.  On the other hand, the high-energy cutoff of the power
law is typically above $100~\unit{keV}$, a significant fraction of the
electron rest mass.  It follows that the optical depth is also of
order unity.  A similar distinction between quasi-thermal and power-law
components is observed in the continuum emission of AGN, except that
the characteristic photon energy of the quasi-thermal component lies in
the rest-rame UV or visible, as predicted by straight-forward scalings
with black-hole mass; the characteristic energy cut-off of the power-law
component remains comparable to the electron rest mass.

It is generally belived that the solar corona is produced and 
maintained by magnetic reconnection events involving coronal loops.
One of us has recently observed that the characteristic plasma density
and magnetic field strength of the solar corona are such that reconnection 
layers should be marginally collisionless, in a sense made precise below, 
and that this may perhaps be a natural outcome of the requirement for fast 
reconnection [\cite{Uzdensky07}; see also \cite{Cassak_etal06}]. 
Similar mechanism have also been shown to work in coronae of other stars
\citep{Cassak_Mullan_Shay08}.
We will show that a modest theoretical extrapolation beyond the framework 
summarized in the previous paragraph leads to the conclusion that disk 
coronae are also marginally collisionless, and that this may explain why 
disk coronae are able both to build up large stores of magnetic free energy 
and to dissipate them efficiently.


\section{Collisional and collisionless reconnection}
\label{sec-recn}

In this section, for the sake of completeness, we rehearse some points 
already made more fully by~\cite{Uzdensky07}. The upshot is the conjecture 
that under astrophysical conditions, fast Petschek-like reconnection 
occurs if and only if the Sweet-Parker thickness (\ref{eq:dSP}) is smaller 
than the ion skin depth (\ref{eq:didef}).

In Sweet-Parker reconnection \citep{Sweet58,Parker57}, reconnecting
field lines are brought together at a speed $V$ related to the
Alfv\'en speed $\VA\equiv B/\sqrt{4\pi\rho}$ by $V\approx\VA S^{-1/2}$, 
where $S\equiv L\VA/\eta$ is the (dimensionless) Lundquist number based 
on the larger dimension of the reconnection layer, $L$, and the plasma 
magnetic diffusivity, $\eta$.  It is presumed that $L$ is a macroscopic 
length characteristic of the size of the system: for example, the radius 
of curvature of a reconnecting magnetic loop.  In most astrophysical 
situations, $S\gg 1$, meaning that the plasma is an excellent conductor 
on macroscopic scales, whence $V/\VA$ is very small.  Also, the theory
predicts that thickness of the layer is
\begin{equation}
  \label{eq:dSP}
\dSP=L S^{-1/2} = \sqrt{\frac{\eta L}{\VA}}\,.
\end{equation}

Sweet-Parker reconnection, though fairly well understood, is far too
slow to explain solar flares, where reconnection is clearly fast in
the sense that $V/\VA\gtrsim 10^{-2}\gg S^{-1/2}$.  The classical
Petschek model was invented to solve this problem \citep{Petschek64}.
In Petschek reconnection, the width~$\delta$ of the reconnection 
region is not very much smaller than its length, $\delta \lesssim 0.1 L$; 
this yields fast reconnection because mass conservation and force 
balance imply $V/\VA\approx \delta/L$.  Recent numerical simulations, 
analytical arguments, and laboratory evidence indicate that in conventional 
one-fluid resistive magnetohydrodynamics (MHD), however, reconnection follows 
the Sweet-Parker rather than the Petschek model [see \citet{Uzdensky07}
and references therein]. By ``conventional'' we mean that the mean 
free path of the current carriers is small compared to all macroscopic 
scales of interest, that $\eta$ is either constant or governed by two-body 
collisions among the carriers, and that  two-fluid effects, such as 
the Hall effect, are unimportant.  It has been suggested that something 
like Sweet-Parker reconnection might proceed much more quickly in a highly 
turbulent, fully three-dimensional MHD because the reconnection layer could 
then be fractal \citep{LazarianVishniac99}, but as yet there is no clear 
evidence for this from laboratory plasmas or simulations.

Something like Petschek's fast reconnection \emph{is} however observed 
when the reconnection layer is collisionless, so that resistive MHD is 
not strictly valid within the layer.  
The relevant condition is typically $\dSP<\di$, where~$\di$ 
is the ion collisionless skin depth, defined by
\begin{equation}
  \label{eq:didef}
  \di\equiv\frac{c}{\omega_{pi}} = \left(
\frac{m_i c^2}{4\pi Z_i^2 e^2 n_i}\right)^{1/2}\,,
\end{equation}
and $\omega_{pi}$ is the ion plasma frequency~\cite[e.g.,][]{Cassak_etal06,
Yamada_etal06,Uzdensky07}. Hereafter we will assume that most of the ions 
are protons, so that $m_i=m_p$ and $Z_i=1$.  The ion skin depth characterizes 
the scale below which inertial effects dominate the response of the ions to 
an electric field.  Since $\dSP$ involves $\eta$, which is normally governed 
by collisions (but see the discussion of inverse-Compton drag below), 
the condition $\dSP<\di$ can be restated in terms of the electron 
mean free path, as $\lambda_{e,\rm mfp}> L (\beta_e m_e/m_i)^{1/2}$, 
where $\beta_e$ is the ratio of the electron pressure within the reconnection 
layer to the magnetic pressure outside it~\citep{Yamada_etal06}. 
This condition highlights the role of plasma collisionality in 
determining the regime of the reconnection process~\citep{Uzdensky07}.


\section{Collisionality of disk coronae}
\label{sec-adc}

In this paper we adopt the popular view that, just as the solar
corona, accretion disk coronae (ADCe) are heated by numerous
reconnection events (flares) involving coronal magnetic
loops~\citep[e.g.,][]{GRV79,Field_Rogers93,diMatteo98,
  Merloni_Fabian02,Uzdensky_Goodman08}.  Following the conjecture
posed by~\S~\ref{sec-recn}, our agenda now is to estimate the ratio
$\dSP/\di$ in accretion disk coronae using the phenomenology
summarized in~\S~\ref{sec-intro}.

It is useful to characterize the accreting system by dimensionless
parameters. We have already introduced the optical depth, $\tau$, 
of the corona.  The electron temperature will be represented by
$\vartheta_e\equiv kT_e/m_e c^2$, and the ratio of the coronal
\emph{ion} pressure to magnetic stress by $\beta_i$.  
The dimensionless accretion rate is $\dot m\equiv \dot M c^2/L_{\rm E}$, 
scaled by the Eddington luminosity $L_{\rm E}= 4\pi GMm_i c/\sigmaT$.
A dimensionless radial coordinate $r\equiv R/\RS$ scaled by the
Schwarzschild radius $\RS=2 GM/c^2$.  The aspect ratio of the corona
is $h\equiv H/R$.  We do not distinguish between the scale heights of
the magnetic stress and of the coronal plasma. The scale height of the 
cooler disk plasma, which is $\sim\dot m R_S$ if the disk is 
radiation-pressure dominated, is presumed to be $\ll H$.  Equilibrium along
coronal field lines requires $k(T_e+T_i)\approx GMm_i H^2/R^3$; this
becomes $(T_i/T_e)+1\approx (m_i/m_e) h^2 r/\vartheta_e$ in our
dimensionless variables.  Since $h$, $r$, and $\vartheta_e$ are 
all thought to be $O(1)$ where the corona is most luminous, 
a strongly two-temperature plasma is indicated: $T_i\gg T_e$.

To estimate the coronal magnetic field, we assume that the ``viscous'' 
torque that balances inward advection of angular momentum in steady state 
is magnetic, and that a fraction $f\le1$ of this torque is transmitted
through the corona. In the sandwich model, the torque is ultimately exerted 
on the disk since the footpoints of the coronal field and most of the
rotational inertia are rooted there.  It seems likely that $f$ is
closer to unity than to zero when the nonthermal power law dominates
the spectrum~\cite{Merloni_Fabian02}.  Then $f\dot M\sqrt{GMR}\approx 
4\pi R^2 H\langle -B_RB_\phi/4\pi\rangle_{\rm corona}$.  
If the opacity due to pairs is not dominant, so that $n_e\approx n_p$, 
this leads to
\begin{equation}
  \label{eq:VA}
\frac{\VA^2}{c^2}\gtrsim f\dot m\tau^{-1} r^{-3/2}
\end{equation}
in dimensionless variables.  We have written this as an inequality in case
the magnetic pitch angle $\tan^{-1}(B_R/B_\phi)\ll 1$, but we expect it to
be an approximate equality.
Next, hydrostatic balance along the field lines yields $\VS^2\approx h^2c^2/r$ 
for the coronal sound speed based on the gas pressure, so
\begin{equation}
  \label{eq:beta}
  \beta\equiv\frac{\VS^2}{\VA^2}\lesssim 
\tau h^2(f\dot m)^{-1} r^{1/2}\,.
\end{equation}
Here $\beta\equiv\beta_i+\beta_e\approx\beta_i$.

We take the magnetic diffusivity at the Spitzer value based on 
the electron temperature, presuming that $T_p\lesssim (m_p/m_e)\, T_e$ 
so that electrons rather than ions carry most of the current:
\begin{equation}
  \label{eq:etaS}
\eta_{\rm S}\approx \vartheta_e^{-3/2}\,c r_e\ln\Lambda\,,
\end{equation}
where $r_e\equiv e^2/m_e c^2$ is the classical radius of the electron.
The relevant electron temperature entering~(\ref{eq:etaS}) is the 
temperature within the Sweet-Parker reconnection layer just before 
the onset of fast reconnection. Importantly, because of the intense 
dissipation in the layer, this temperature may be as high as that 
obtained during the subsequent fast reconnection phase:
while the speed at which the field lines approach one another is 
very different for the two modes of reconnection, the magnetic 
energy dissipated per unit mass of plasma entering the region 
is approximately the same. Furthermore, because of the rapid 
Compton cooling of the coronal electrons following a fast reconnection 
event, this reconnection-layer temperature can be much higher than the 
ambient coronal electron temperature. Thus, the covering fraction of 
the hot ($T_e\sim 100\unit{keV}$) corona may be very small, consistent 
with observations~\citep{Haardt_etal94,Stern_etal95,Nandra_Papadakis01}.  

Alternatively, at high luminosities the mobility of the electrons 
can be dominated by Compton drag rather than Coulomb collisions. 
If Coulomb collisions with the ions could be entirely neglected, 
then the drift velocity of the (non-relativistic) electrons with 
respect the rest frame of the radiation field would become, in 
steady state, $\boldsymbol{v}_e= -(3/4)\,(ce/\sigmaT U_{\rm rad})
\boldsymbol{E}$,
where $U_{\rm rad}$ is the energy density in the radiation field 
and $\boldsymbol{E}$ is the electric field in that frame. 
The contribution of the electrons to the conductivity would 
then become $\sigma_{\rm C}= (3/4)\,ce^2 n_e/\sigmaT U_{\rm rad}$
(if the electrons are magnetized, then the conductivity becomes 
a tensor, but this formula for $\sigma_{\rm C}$ 
still relates the component of the electron 
current parallel to the magnetic field to the parallel component 
of $\boldsymbol{E}$).  Expressed in our dimensionless variables, 
the corresponding resistivity $\eta_{\rm C}=c^2/4\pi\sigma_{\rm C}$
would then be
\begin{equation}
  \label{eq:etaC}
  \eta_{\rm C}\approx \frac{c\sigmaT U_{\rm rad}}{3\pi e^2 n_e} \sim
\frac{m_p}{m_e}\frac{\epsilon\dot m h}{r}\,cr_e\,,
\end{equation}
where $\epsilon\equiv L/\dot M c^2$ is the radiative efficiency 
of accretion, and we have taken the photon energy density to be 
$U_{\rm rad} \sim\tau L/4\pi R^2 c \sim (\epsilon\dot{m}\tau/r^2) \,
(m_p c^2/R_S\sigma_T) $ with $\tau\gtrsim 1$.
A relativistic generalization of the first form of eq.~(\ref{eq:etaC}) 
has been given by~\cite{vanOss_etal93}.

Since each positron contributes in the same way as an electron to
the conductivity and to the optical depth, the estimate~(\ref{eq:etaC}) 
for Compton diffusivity is unaffected by the presence of pairs, at least 
to the extent that their density in the corona is approximately uniform---but
note that it may be enhanced in reconnection regions.  We should
consider whether the ions 
may dominate the conductivity, despite their greater inertia, when the 
leptons are immobilized by Compton drag.  For an infinite and homogeneous 
electron-ion plasma, in fact, the DC (i.e., static) resistivity would be 
given by the Spitzer formula (\ref{eq:etaS}) regardless of the radiation
field.  This can be seen by considering the forces on the electrons and 
ions.  Electric fields exert equal and opposite forces on the two charged 
species, whereas to a good approximation, radiation drag acts only on the 
electrons.  Therefore in steady state, charge neutrality and force balance 
require the net Compton drag on the electrons to vanish.%
\footnote{ We are ignoring gravity, which acts mainly on the ions.
  The electric field required to keep the electrons and ions from
  separating gravitationally is a potential field, whereas the
  electric fields associated with time derivatives of magnetic flux
  have nonzero curl. Potential fields do not affect the present
  discussion.}  Hence the mean velocity of the electrons should vanish
in the rest frame of the radiation field.  If momentum exchange
between the electrons and ions is due solely to Coulomb encounters,
then for a given electric field, the same drift velocity between the
two species results as if the radiation field were absent, and hence
the same conductivity.  However, the ions probably cannot reach steady
state during reconnection in an actual disk corona.  While the
electron mean free path corresponding to (\ref{eq:etaS}) is
substantially smaller than the coronal scale height, $\lambda_{\rm
  m.f.p.}/H \approx \vartheta_e^2/(\tau\ln\Lambda)\lesssim 10^{-2}$
for standard parameters, the collisional mean free path of the protons
is larger by the factor $m_p/m_e$ and hence is $>H$.  Thus the ions
are not expected to reach their full drift velocity before leaving the
reconnection layer, whose length is presumably $<H$; the ions
decelerate only when they collide with the disk.  So the contribution
of the ions to the current is probably less than that of the
electrons.

In short, we expect that the effective diffusivity for
collisional reconnection is approximately the larger of $\eta_{\rm S}$
and $\eta_{\rm C}$.
Comparing equations (\ref{eq:etaS}) and (\ref{eq:etaC}), we see that
the ratio of Compton to Spitzer resistivities is roughly
\begin{equation}
\frac{\eta_C}{\eta_S} \sim  
\frac{\theta_e^{3/2}}{\ln\Lambda}\, \frac{U_{\rm rad}}{n m_e c^2} \sim 
\frac{1}{\ln\Lambda}\,\frac{m_p}{m_e}\, \dot{m}\, \epsilon\, 
h\,\theta_e^{3/2}\, r^{-1}\,.
\end{equation}
With $\ln\Lambda\sim 20$, $\vartheta_e\sim 0.2$, and $\epsilon\sim 0.1$, 
the contributions (\ref{eq:etaS}) and~(\ref{eq:etaC}) to the total 
resistivity may often be comparable near the inner edge of the disk, 
but the Spitzer resistivity probably dominates at larger radii because
$\eta_{\rm S}/\eta_{\rm C}\propto r\vartheta_e^{-3/2}$.

If one takes the scale height $H=hr\RS$ as the characteristic macroscopic
length, the coronal Lundquist number based on the Spitzer resistivity becomes
\begin{equation}
  \label{eq:Scorona}
  S_{\rm S}= \frac{H\VA}{\eta_S}\sim \left(\frac{\RS}{r_e\ln\Lambda}\right)
f^{1/2}\dot m^{1/2}\tau^{-1/2}\vartheta_e^{3/2} h r^{1/4}\,.
\end{equation}
Similarly, the Lundquist number based on the Compton resistivity is
\begin{equation}
  \label{eq:Scorona-Compton}
  S_{\rm C}= \frac{H\VA}{\eta_C}\sim \left(\frac{\RS}{r_e}\right)
\frac{m_e}{m_p}\, f^{1/2}\dot{m}^{-1/2}\tau^{-1/2} \epsilon^{-1} r^{5/4}\,.
\end{equation}
All of the dimensionless factors except the first are expected to be
of order unity, but $\RS/(r_e\ln\Lambda)\sim 10^{17}(M/M_\odot)$ with
$\ln\Lambda\approx 20$.  The coronae of accreting black holes are therefore
very, very deeply within the ideal-MHD regime, and
Sweet-Parker reconnection is completely negligible.
The ion skin depth $\di=(m_p c^2/4\pi e^2 n_i)^{1/2}$ becomes
\begin{equation}
  \label{eq:di}
  \di\approx \left(\frac{m_p}{m_e}\frac{hr}{\tau}\right)^{1/2} 
  \sqrt{r_e\RS} 
\end{equation}
if we take $n_i\approx n_e$ as before.  This shows that the
characteristic scale for fast reconnection---the thickness of the
current sheet---is approximately the geometric mean between two
fundamental physical scales relevant to electromagnetic processes near
a black hole: the size of the black hole and the classical electron
radius.

Thus, we have the following expression for the collisionality parameter,
\begin{eqnarray}
  \label{eq:ratio}
\frac{\dSP}{\di} & \sim &
\sqrt{\frac{m_e}{m_p}\,\ln\Lambda} \, 
f^{-1/4}\dot m^{-1/4}\tau^{3/4}\vartheta_e^{-3/4}r^{3/8}\, , \ 
\quad\mbox{if } \eta=\eta_S \, , \nonumber \\
\frac{\dSP}{\di} & \sim &
\epsilon^{1/2}f^{-1/4}\dot{m}^{1/4}\tau^{3/4} h^{1/2} r^{-1/8} \, , \
\quad\mbox{if } \eta=\eta_C \, .  
\end{eqnarray}
Recall that reconnection is conjectured to be very slow when this
ratio is $>1$ (the collisional, conventional MHD regime) but can be
fast (reconnecting field lines approaching at speeds $\sim\VA$ rather
than $\sim\VA S^{-1/2}$) when the ratio is less than unity.


\section{Discussion}

Marginal collisionality may explain why the particular combination 
of dimensionless parameters that enters~eq.~(\ref{eq:ratio}) should 
be of order unity, but it does not explain why the individual factors 
are also of order unity: in particular, the coronal optical depth ($\tau$) 
and dimensionless temperature ($\vartheta_e$).  
Additional constraints on these parameters have long been recognized.  
The combinations $\epsilon\dot m$, $f$, and $r^{-1}$ should be $< 1$ for 
physical reasons, and it is not surprising that sources selected for high 
luminosity and prominent nonthermal emission approach these limits.  
It is plausible that relativistic corrections to Compton scattering 
and pair production explain why $\vartheta_e\lesssim O(1)$.  
In conjunction with these more familiar constraints, the condition 
$\delta_{\rm SP}\sim d_i$ then ``explains'' why $\tau\sim O(1)$. 
On the other hand, one might argue that there are other reasons 
why $\tau\sim O(1)$: 
for example, the Thompson optical depth is about unity near $r=1$ 
in spherical accretion at the free-fall speed with $\dot m=1$
\cite[e.g.,][]{Rees84}. On this view, our result $\delta_{\rm SP}\sim d_i$
is merely a numerical coincidence. 
At the very least, however, the considerations of this paper 
indicate that reconnection in ADC is likely to occur within the 
marginally collisionless regime.

Taken together with equations (\ref{eq:VA}) and~(\ref{eq:di}), 
$\tau\sim O(1)$ implies that inner parts of ADCe of accreting black 
holes accreting near Eddington limit are characterized by the following 
fiducial values for plasma density, magnetic field strength, and the fast 
reconnection scale, which only depend on the black hole mass:
\begin{eqnarray}
n_* &=& (\sigma_T R_g)^{-1} \simeq 10^{19}\, {\rm cm}^{-3}\, m^{-1}\,,
\label{eq:fiducial-n}  \\
B_* &=& c\,\sqrt{4\pi\,n_* m_p} \simeq 4\cdot 10^8\, m^{-1/2}\, {\rm G} \,, 
\label{eq:fiducial-B} \\
d_{i*} &=& (4\pi n_* r_i)^{-1/2} \simeq  10^{-2}\, m^{1/2} {\rm cm} \, ,
\label{eq:fiducial-d}
\end{eqnarray}
where $m\equiv M/M_\odot$.

Most of the observational evidence for ADC pertains to the innermost 
part of the disk, where the physical parameters are comparable to those in
equations~(\ref{eq:fiducial-n})--(\ref{eq:fiducial-d}). However, the outer 
parts may also have coronae. The hypothesis of marginal collisionality 
would then constrain the radial dependence of coronal properties. 
Equation (\ref{eq:ratio}) leads one to expect that $\tau$ or 
$\tau/\vartheta_e$ should vary as~$r^{-1/2}$ in a given source.  
(In the second alternative on the right-hand side of that equation, 
$\tau^{1/4}$ replaces $\tau^{3/4}$ where $\tau<1$ because the radiation 
energy density becomes independent of~$\tau$.)  
It would be interesting to explore the consequences of this for eclipsing 
systems such as the white dwarf binary OY~Car \citep{Wood_etal89}, for 
example, where it may be possible to constrain the radial extent and 
structure of the X-ray-emitting region~\citep{Naylor_etal88,Ramsay_etal01};
we expect that disk coronae should not be limited to black-hole systems.

The train of thought in this paper calls attention to some important
issues. For example, how does magnetic reconnection proceed with
marginally relativistic electrons and much hotter, perhaps also 
marginally relativistic, ions? How is it affected by the presence 
of a strong radiation background? What is the effect 
of pair creation on the reconnection process? 
What fraction of the magnetic energy released in the reconnection process 
and associated shocks goes to thermal electrons and nonthermal electrons? 
Does an appreciable fraction of the dissipated energy go into the ions, 
as appears to be the case for flares in the solar corona~\citep[e.g.,][]
{Ramaty_etal95,Lin_etal03,Benz08}? What is the fate of the energy
that goes to the ions? Whereas the main cooling mechanism for coronal
electrons is probably Comptonization,
the ions must cool (if at all) by other means.
Some of their energy may be advected to infinity or into the event 
horizon in winds or plasmoids (i.e., disconnected loops 
of field and plasma).  But the hot ions on closed field lines that are still 
tied to the slowly accreting disk seem likely to cool by collisions 
with the disk itself, unless this is inhibited by magnetic mirroring. 
Thus, \emph{ion} thermal conduction, in addition to the direct irradiation 
by hard X-ray coronal emission, can be an imporant channel for transporting 
the energy dissipated in the corona down to the cool accretion disk. 
An important question then is, what are the main stopping mechanisms 
for hot coronal ions entering the much colder disk plasma and what are 
their observable signatures?

We leave these interesting questions for future study, but it seems
necessary to address a more central issue here: What are the feedback
mechanisms that maintain the corona in a state of marginal collisionality?  
For the solar corona, \cite{Uzdensky07} proposed that the answer lies 
in the effect of reconnection on the balance between coronal heating 
and cooling.  If the plasma density in the corona becomes large enough 
so that $\delta_{\rm SP}>d_i$ (i.e., the collisional regime), then 
reconnection becomes extremely slow and the rate of heating by magnetic 
energy dissipation falls sharply; the plasma then cools and drains down 
the field lines onto the photosphere until marginal collisionality is 
restored.  Conversely, as the plasma density drops, fast reconnection 
converts magnetic energy into X-rays and energetic particles that evaporate 
photospheric gas into the corona.  Should the coronal density nevertheless 
fall to an extremely low value and the coronal magnetic field become 
potential (i.e., current-free: $\boldsymbol{\nabla\times B}=0$)
through efficient reconnection, the field will offer less 
resistance to plasma-laden flux loops emerging from the turbulent 
photosphere. We expect that feedback mechanisms similar to these should 
operate in accretion disk coronae, but there may be some new effects due 
to the high luminosity and relativistic conditions. For example, 
pair creation inside the reconnection region may substantially increase 
the plasma density, thereby making the layer more collisional and inhibiting
fast reconnection.

Finally, it is interesting to note that the coronal scale-height cancels 
when the Spitzer resistivity dominates.  In fact, the first alternative 
in equation~(\ref{eq:ratio}) would apply to the disk itself if 
the parameters $(f,\tau,\vartheta_e)$ were replaced by the torque 
fraction, optical depth, and dimensionless electron temperature of 
the disk. We expect $\tau_{\rm disk}\gtrsim 1$ and
$\vartheta_{e,\rm disk}\ll 1$, so the disk is probably collisional
and therefore, we conjecture, incapable of fast reconnection.  
The conclusion is stronger for more massive accretors (\emph{i.e.}, 
AGN rather than X-ray binaries) because the characteristic disk
temperature at a fixed Eddington fraction $\dot m$ varies as
$M^{-1/4}$, whereas the optical depth of the disk is independent of
$M$ for a given $\dot m$ and Shakura-Sunyaev $\alpha$ parameter.  
Yet even AGN show powerlaw X-ray spectra that suggest
$\vartheta_{e}\gtrsim 0.1$, so they probably have active coronae.

\begin{acknowledgments}

We would like to acknowledge fruitful discussions with A.~Socrates.

This work is supported by National Science Foundation Grant 
No.\, PHY-0215581 (PFC: Center for Magnetic Self-Organization 
in Laboratory and Astrophysical Plasmas).

\end{acknowledgments}


\begin{thebibliography}{30}
\expandafter\ifx\csname natexlab\endcsname\relax\def\natexlab#1{#1}\fi

\bibitem[{{Band} \& {Grindlay}(1986)}]{Band_Grindlay86}
{Band}, D.~L. \& {Grindlay}, J.~E. 1986, \apj, 308, 576

\bibitem[{{Benz}(2008)}]{Benz08}
{Benz}, A.~O. 2008, Living Reviews in Solar Physics, 5, 1

\bibitem[{{Cassak} {et~al.}(2006){Cassak}, {Drake}, \& {Shay}}]{Cassak_etal06}
{Cassak}, P.~A., {Drake}, J.~F., \& {Shay}, M.~A. 2006, \apjl, 644, L145

\bibitem[{{Cassak} {et~al.}(2008){Cassak}, {Mullan}, \&
  {Shay}}]{Cassak_Mullan_Shay08}
{Cassak}, P.~A., {Mullan}, D.~J., \& {Shay}, M.~A. 2008, \apjl, 676, L69

\bibitem[{{di Matteo}(1998)}]{diMatteo98}
{di Matteo}, T. 1998, \mnras, 299, L15+

\bibitem[{{Eardley} \& {Lightman}(1976)}]{Eardley_Lightman76}
{Eardley}, D.~M. \& {Lightman}, A.~P. 1976, \nat, 262, 196

\bibitem[{{Field} \& {Rogers}(1993)}]{Field_Rogers93}
{Field}, G.~B. \& {Rogers}, R.~D. 1993, \apj, 403, 94

\bibitem[{{Galeev} {et~al.}(1979){Galeev}, {Rosner}, \& {Vaiana}}]{GRV79}
{Galeev}, A.~A., {Rosner}, R., \& {Vaiana}, G.~S. 1979, \apj, 229, 318

\bibitem[{{Haardt} {et~al.}(1994){Haardt}, {Maraschi}, \&
  {Ghisellini}}]{Haardt_etal94}
{Haardt}, F., {Maraschi}, L., \& {Ghisellini}, G. 1994, \apjl, 432, L95

\bibitem[{{Lazarian} \& {Vishniac}(1999)}]{LazarianVishniac99}
{Lazarian}, A. \& {Vishniac}, E.~T. 1999, \apj, 517, 700

\bibitem[{{Liang} \& {Price}(1977)}]{Liang_Price77}
{Liang}, E.~P.~T. \& {Price}, R.~H. 1977, \apj, 218, 247

\bibitem[{{Lin} {et~al.}(2003){Lin}, {Krucker}, {Hurford}, {Smith}, {Hudson},
  {Holman}, {Schwartz}, {Dennis}, {Share}, {Murphy}, {Emslie}, {Johns-Krull},
  \& {Vilmer}}]{Lin_etal03}
{Lin}, R.~P., {Krucker}, S., {Hurford}, G.~J., {Smith}, D.~M., {Hudson}, H.~S.,
  {Holman}, G.~D., {Schwartz}, R.~A., {Dennis}, B.~R., {Share}, G.~H.,
  {Murphy}, R.~J., {Emslie}, A.~G., {Johns-Krull}, C., \& {Vilmer}, N. 2003,
  \apjl, 595, L69

\bibitem[{{Merloni} \& {Fabian}(2002)}]{Merloni_Fabian02}
{Merloni}, A. \& {Fabian}, A.~C. 2002, \mnras, 332, 165

\bibitem[{{Nandra} \& {Papadakis}(2001)}]{Nandra_Papadakis01}
{Nandra}, K. \& {Papadakis}, I.~E. 2001, \apj, 554, 710

\bibitem[{{Naylor} {et~al.}(1988){Naylor}, {Bath}, {Charles}, {Hassal},
  {Sonneborn}, {van der Woerd}, \& {van Paradijs}}]{Naylor_etal88}
{Naylor}, T., {Bath}, G.~T., {Charles}, P.~A., {Hassal}, B.~J.~M., {Sonneborn},
  G., {van der Woerd}, H., \& {van Paradijs}, J. 1988, \mnras, 231, 237

\bibitem[{{Parker}(1957)}]{Parker57}
{Parker}, E.~N. 1957, \jgr, 62, 509

\bibitem[{{Petschek}(1964)}]{Petschek64}
{Petschek}, H.~E. 1964, in The Physics of Solar Flares, ed. W.~N. {Hess},
  425--+

\bibitem[{{Ramaty} {et~al.}(1995){Ramaty}, {Mandzhavidze}, {Kozlovsky}, \&
  {Murphy}}]{Ramaty_etal95}
{Ramaty}, R., {Mandzhavidze}, N., {Kozlovsky}, B., \& {Murphy}, R.~J. 1995,
  \apjl, 455, L193+

\bibitem[{{Ramsay} {et~al.}(2001){Ramsay}, {Poole}, {Mason}, {C{\'o}rdova},
  {Priedhorsky}, {Breeveld}, {Much}, {Osborne}, {Pandel}, {Potter}, {West}, \&
  {Wheatley}}]{Ramsay_etal01}
{Ramsay}, G., {Poole}, T., {Mason}, K., {C{\'o}rdova}, F., {Priedhorsky}, W.,
  {Breeveld}, A., {Much}, R., {Osborne}, J., {Pandel}, D., {Potter}, S.,
  {West}, J., \& {Wheatley}, P. 2001, \aap, 365, L288

\bibitem[{{Rees}(1984)}]{Rees84}
{Rees}, M.~J. 1984, \araa, 22, 471

\bibitem[{{Shapiro} {et~al.}(1976){Shapiro}, {Lightman}, \&
  {Eardley}}]{Shapiro_etal76}
{Shapiro}, S.~L., {Lightman}, A.~P., \& {Eardley}, D.~M. 1976, \apj, 204, 187

\bibitem[{{Stern} {et~al.}(1995){Stern}, {Poutanen}, {Svensson}, {Sikora}, \&
  {Begelman}}]{Stern_etal95}
{Stern}, B.~E., {Poutanen}, J., {Svensson}, R., {Sikora}, M., \& {Begelman},
  M.~C. 1995, \apjl, 449, L13+

\bibitem[{{Sweet}(1958)}]{Sweet58}
{Sweet}, P.~A. 1958, in IAU Symposium, Vol.~6, Electromagnetic Phenomena in
  Cosmical Physics, ed. B.~{Lehnert}, 123

\bibitem[{{Thorne} \& {Price}(1975)}]{Thorne_Price75}
{Thorne}, K.~S. \& {Price}, R.~H. 1975, \apjl, 195, L101

\bibitem[{{Uzdensky} \& {Goodman}(2008)}]{Uzdensky_Goodman08}
{Uzdensky}, D. \& {Goodman}, J. 2008, accepted to \apj; ArXiv e-print:0803.0337

\bibitem[{{Uzdensky}(2007)}]{Uzdensky07}
{Uzdensky}, D.~A. 2007, \apj, 671, 2139

\bibitem[{{van Oss} {et~al.}(1993){van Oss}, {van den Oord}, \&
  {Kuperus}}]{vanOss_etal93}
{van Oss}, R.~F., {van den Oord}, G.~H.~J., \& {Kuperus}, M. 1993, \aap, 270,
  275

\bibitem[{{Wood} {et~al.}(1989){Wood}, {Horne}, {Berriman}, \&
  {Wade}}]{Wood_etal89}
{Wood}, J.~H., {Horne}, K., {Berriman}, G., \& {Wade}, R.~A. 1989, \apj, 341,
  974

\bibitem[{{Yamada} {et~al.}(2006){Yamada}, {Ren}, {Ji}, {Breslau}, {Gerhardt},
  {Kulsrud}, \& {Kuritsyn}}]{Yamada_etal06}
{Yamada}, M., {Ren}, Y., {Ji}, H., {Breslau}, J., {Gerhardt}, S., {Kulsrud},
  R., \& {Kuritsyn}, A. 2006, Physics of Plasmas, 13, 2119

\bibitem[{{Zdziarski} \& {Gierli{\'n}ski}(2004)}]{Zdziarski_Gierlinski04}
{Zdziarski}, A.~A. \& {Gierli{\'n}ski}, M. 2004, Progress of Theoretical
  Physics Supplement, 155, 99

\end{thebibliography}

\end{document}